\documentclass[conference, 10pt]{IEEEtran}

\usepackage[english]{babel}
\usepackage[T1]{fontenc}
\usepackage[utf8]{inputenc}

\usepackage{amsmath,amssymb}
\usepackage{xcolor}
\usepackage{colortbl}
\usepackage{booktabs}
\usepackage{graphicx}
\usepackage{tabularx}

\usepackage{placeins}
\usepackage{float}

\usepackage{pgfplots}
\pgfplotsset{compat=newest}
\pgfplotsset{plot coordinates/math parser=false}
\newlength\figureheight
\newlength\figurewidth 

\graphicspath{{images/}}

\begin{document}
%
\title{Texture-Dependent Frequency Selective Reconstruction of Non-Regularly Sampled Images}

\author{\IEEEauthorblockN{Markus Jonscher, Jürgen Seiler, and André Kaup}
\IEEEauthorblockA{Multimedia Communications and Signal Processing \\ Friedrich-Alexander University Erlangen-Nürnberg (FAU), Cauerstr. 7, 91058 Erlangen, Germany \\ \{markus.jonscher, juergen.seiler, andre.kaup\}@fau.de}
}


%


\maketitle

\begin{abstract}
\boldmath
There exist many scenarios where pixel information is available only on a non-regular subset of pixel positions. For further processing, however, it is required to reconstruct such images on a regular grid. Besides many other algorithms, frequency selective reconstruction can be applied for this task. It performs a block-wise generation of a sparse signal model as an iterative superposition of Fourier basis functions and uses this model to replace missing or corrupted pixels in an image. In this paper, it is shown that it is not required to spend the same amount of iterations on both homogeneous and heterogeneous regions. Hence, a new texture-dependent approach for frequency selective reconstruction is introduced that distributes the number of iterations depending on the texture of the regions to be reconstructed. 
Compared to the original frequency selective reconstruction and depending on the number of iterations, visually noticeable gains in PSNR of up to $1.47$~dB  can be achieved.
\end{abstract}

%
%

%
\IEEEpeerreviewmaketitle

\section{Introduction}
\label{sec:intro}
\begin{figure}
	\centering
	\def\svgwidth{\columnwidth}
	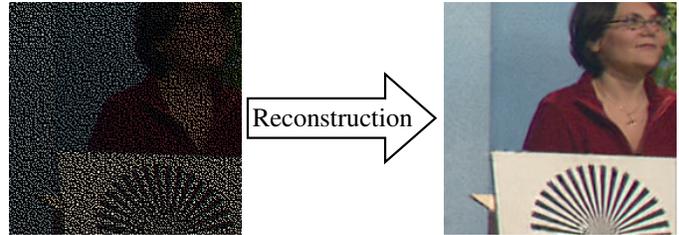
	\caption{Example for the reconstruction of non-regularly sampled image data on a regular grid.}
	\label{fig:motivation_image_reconstruction}
\end{figure}

Digital images are typically defined on a regular grid which is in many cases caused by the acquisition systems. A regular grid is in most cases required for a further processing of these images and also for displaying them. There are, however, many scenarios where pixel information is available only on a non-regular subset of pixel positions. This might be caused for example by a specific acquisition system~\cite{Duparre2006} or intentionally in order to reduce the influence of aliasing~\cite{Hennenfent2007}. Another example can be found in~\cite{Schoeberl2011} where the spatial resolution of an imaging sensor is increased by a non-regular masking of the sensor.
In Figure~\ref{fig:motivation_image_reconstruction},  an example for the task of reconstructing non-regularly sampled image data on  a regular grid is shown. On the left, an image that is taken by a non-regular sampling sensor can be seen. The idea behind such a non-regular sampling sensor is that a low resolution sensor can be used in order to obtain a high resolution image~\cite{Schoeberl2011}. Therefore, each pixel of the low resolution sensor is divided into four parts where three of them are randomly covered. In doing so, an image with twice the resolution in both spatial dimensions is acquired. Due to the masking, however, $75\%$ of the pixels are missing and have to be reconstructed by a suitable reconstruction algorithm in order to obtain the full high resolution image. Besides frequency selective reconstruction (FSR)~\cite{Seiler2015}, there are many other reconstruction algorithms like linear interpolation (LIN), steering kernel regression (SKR)~\cite{Takeda2007}, the constrained split augmented Lagrangian shrinkage algorithm (CLS)~\cite{Afonso2011}, or sparsity-based wavelet inpainting (WI)~\cite{Starck2010} that can be used for such reconstruction tasks. However, it has been shown in~\cite{Seiler2015} that for a non-regular subsampling problem like this, FSR yields a better reconstruction quality than the other mentioned state-of-the-art image reconstruction algorithms.
The basic idea behind FSR is the block-wise generation of a sparse signal model as an iterative superposition of Fourier basis functions, where the number of iterations is fixed for each block. Missing or corrupted pixels are then replaced by utilizing the generated model.
The advantage of a non-regular subsampling is that compared with a regular subsampling, dominant basis functions can still be identified in the available non-regularly subsampled signal. A detailed discussion of this topic can be found in~\cite{Seiler2015}.

In this paper, the number of iterations per block used by FSR is adapted to the texture of the block to be reconstructed. Since homogeneous blocks need only fewer iterations and heterogeneous blocks which contain more structure need more iterations, firstly, an image segmentation is required in order to divide the image into homogeneous and heterogeneous areas. By applying a linear mapping to this segmentation, the resulting number of iterations per block can be determined. Hence, the same number of iterations that would be employed by the original FSR can now be distributed in a more sophisticated way over all blocks. By using the new texture-dependent FSR, an improved image reconstruction quality is achievable especially in scenarios where only few iterations can be spent which might be the case for example in mobile devices with low computational power.

The paper is organized as follows: The next section introduces the state-of-the-art image reconstruction algorithm FSR and Section~\ref{sec:proposed} presents the proposed texture-dependent FSR. Comprehensive simulations and results are given in Section~\ref{sec:results} and Section~\ref{sec:conclusion} concludes this contribution.

\section{Frequency Selective Reconstruction}
\label{sec:fsr}
In this section, frequency selective reconstruction (FSR) \cite{Seiler2015} is introduced. It is a general algorithm for signal reconstruction and it is based on frequency selective extrapolation~\cite{Seiler2010}. It has been used for example in error concealment scenarios~\cite{Koloda2014} or the reconstruction of non-regularly sampled video data~\cite{Jonscher2016}. 
\begin{figure}
	\hspace*{0.25cm}
	\def\svgwidth{0.85\columnwidth}
	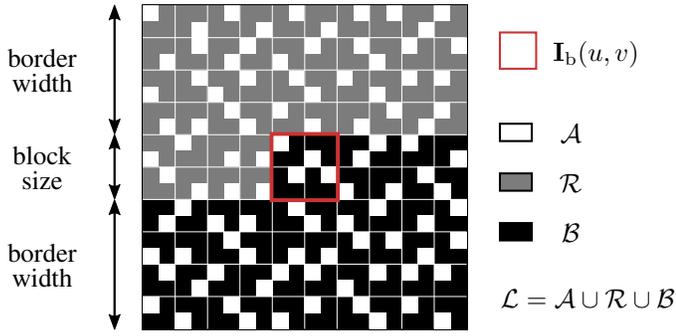
	\caption{Reconstruction area $\mathcal{L}$ consisting of available pixels in support area $\mathcal{A}$, missing pixels in loss area $\mathcal{B}$, and already reconstructed pixels in area $\mathcal{R}$.}
	\label{fig:FSR_extrapolation_area}
\end{figure}
FSR performs a block-wise reconstruction and Figure~\ref{fig:FSR_extrapolation_area} shows the reconstruction area 
\begin{equation}
\mathcal{L} = \mathcal{A}\cup\mathcal{R}\cup\mathcal{B}
\end{equation}
which consists of the block $\mathbf{I}_{\mathrm{b}}(u,v)$ that is currently to be reconstructed (red frame, block size) and a certain surrounding (border width). 
The reconstruction area $\mathcal{L}$ is depicted by the coordinates $(m,n)$ and is of size $M\times N$. $(u,v)$ depict the coordinates of the block that is currently to be reconstructed which is of size $U\times V$.
The reconstruction area $\mathcal{L}$ is divided into three areas, where all originally acquired pixels are located in the support area $\mathcal{A}$ and all already reconstructed pixels in area $\mathcal{R}$. Area $\mathcal{B}$ contains all missing pixels that have to be reconstructed.

The basic idea of FSR is to generate the sparse signal model
\begin{equation}
	g[m,n] = \sum\limits_{(k,l)\in\mathcal{K}}\hat{c}_{(k,l)}\varphi_{(k,l)}[m,n]
\end{equation}
as a superposition of Fourier basis functions $\varphi_{(k,l)}[m,n]$ weighted by the expansion coefficients $\hat{c}_{(k,l)}$. The set $\mathcal{K}$ contains the indices $(k,l)$ of all basis functions that have been used for model generation. In every iteration, one basis function gets selected and before being added to the model, its corresponding weight is estimated.
Additionally, the weighting function
\begin{equation}
	w[m,n] = 	\begin{cases}
					\hat{\rho}^{\sqrt{\left(m-\frac{M-1}{2}\right)^2+\left(n-\frac{N-1}{2}\right)^2}} & \forall \left(m,n\right)\in\mathcal{A} \\
					\delta\hat{\rho}^{\sqrt{\left(m-\frac{M-1}{2}\right)^2+\left(n-\frac{N-1}{2}\right)^2}} & \forall \left(m,n\right)\in\mathcal{R} \\
					0 & \forall \left(m,n\right)\in\mathcal{B}
				\end{cases}
\end{equation}
is used to control the influence that each pixel has on the model generation. Pixels in the support area $\mathcal{A}$ that are farther away from $\mathbf{I}_{\mathrm{b}}(u,v)$ get less weight than pixels closer to it. $\hat{\rho}$ controls this decay. Since pixels in area $\mathcal{R}$ are not as reliable as pixels in area $\mathcal{A}$, they have an additional attenuation factor $\delta$. Pixels in the loss area $\mathcal{B}$ are neglected for model generation and therefore set to zero.
After a fixed number of iterations, $g[m, n]$ is used to replace missing pixels in the corrupted image.


\section{Proposed Texture-Dependent Frequency Selective Reconstruction}
\label{sec:proposed}
\begin{figure}
	\centering
%
%
\definecolor{mycolor1}{rgb}{0.00000,0.44700,0.74100}%
\begin{tikzpicture}

\begin{axis}[%
width=0.85\columnwidth,
height=0.543\columnwidth,
at={(0\columnwidth,0\columnwidth)},
scale only axis,
xmin=0,
xmax=80,
xlabel={Iterations $i_{\mathrm{b,f}}$},
xmajorgrids,
ymin=10,
ymax=50,
ylabel={PSNR [dB]},
ymajorgrids,
axis background/.style={fill=white},
legend style={at={(0.97,0.03)},anchor=south east,legend cell align=left,align=left,draw=white!15!black}
]
\addplot [color=mycolor1,solid,mark size=2.0pt,mark=*,mark options={solid,fill=white}]
  table[row sep=crcr]{%
2	17.8585501885087\\
4	29.1862150718797\\
6	39.9240692018507\\
8	46.01184271206\\
10	46.4961149437107\\
20	46.6622587760144\\
40	46.247232305533\\
60	46.1493896106915\\
80	45.9910630321346\\
100	45.7786764969357\\
};
\addlegendentry{FSR on homogeneous block};

\addplot [color=mycolor1,dashdotted,mark size=2.5pt,mark=triangle*,mark options={solid,fill=white}]
  table[row sep=crcr]{%
2	14.5731615163331\\
4	22.0336509027553\\
6	23.9664139587109\\
8	27.8140370160286\\
10	29.2283841366628\\
20	35.7698022735344\\
40	37.4506853017273\\
60	37.7852710023999\\
80	38.2897517081856\\
100	37.8293621914505\\
};
\addlegendentry{FSR on heterogeneous block};

\end{axis}
\end{tikzpicture}%
	\vspace*{0.03cm}
	\def\svgwidth{1.0\columnwidth}
	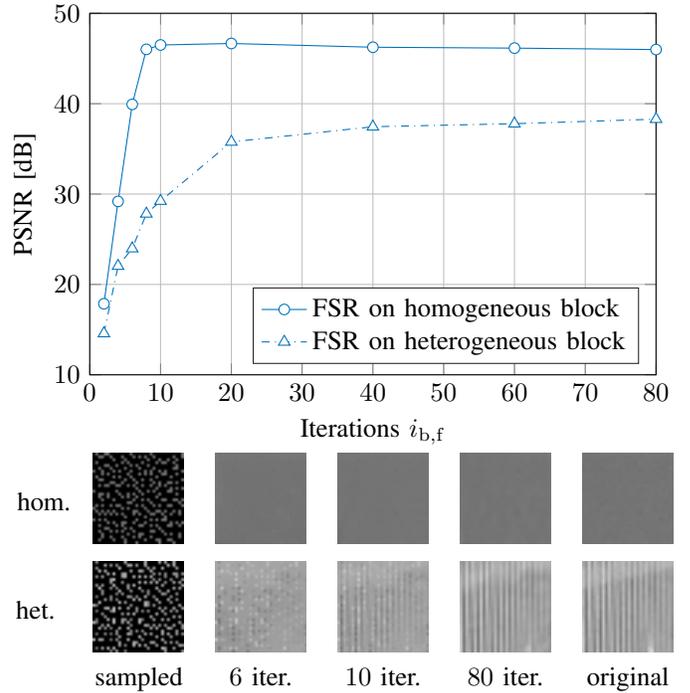
	\caption{Top: Number of iterations needed for a proper reconstruction quality of a homogeneous and heterogeneous block. Bottom: A homogeneous and heterogeneous block of \textit{Kodim19} (KODAK) reconstructed using different iterations.}
	\label{fig:psnr_results_FSR_different_blocks}
\end{figure}

FSR requires a certain number of iterations $i_\mathrm{b}$ for each block $\mathbf{I}_{\mathrm{b}}(u,v)$ in order to generate an adequate signal model. Up to now, FSR uses a fixed number of iterations $i_{\mathrm{b,f}}$ for each block. This leads to a total amount of iterations
\begin{equation}
	i_{\mathrm{t,f}} = \sum\limits_{b=1}^{B}i_{\mathrm{b,f}}
\end{equation}
for the entire image, where $B$ is the number of blocks that have to be reconstructed.
As it can be seen from the upper part of Figure~\ref{fig:psnr_results_FSR_different_blocks}, homogeneous blocks do not need as many iterations as heterogeneous blocks in order to obtain a good reconstruction quality. By choosing \mbox{$i_{\mathrm{b,f}} = 10$} for example, the reconstruction quality of homogeneous blocks already has reached its maximum, whereas for heterogeneous blocks, more iterations are required. Image detail examples for the reconstruction of homogeneous and heterogeneous blocks are given in the lower part of Figure~\ref{fig:psnr_results_FSR_different_blocks}.
It can be seen that for a homogeneous area like sky, very few iterations are sufficient to obtain a satisfactory visual quality. For heterogeneous areas like fences, however, many more iterations have to be spent to get visually pleasing results.
As a consequence, FSR should spend less iterations on homogeneous blocks and more iterations on heterogeneous blocks. FSR using such variable iterations is from now on referred to as texture-dependent frequency selective reconstruction~\mbox{(TD-FSR)}. To adapt the iterations $i_{\mathrm{b}}$ to the texture of the block to be reconstructed, a segmentation is required. Therefore, for each block $\mathbf{I}_{\mathrm{b}}(u,v)$, the variance
\begin{equation}
	\sigma^2_{\mathrm{b}} = \frac{1}{U-1}\frac{1}{V-1}\sum\limits_{u=1}^{U}\sum\limits_{v=1}^{V}\left|\mathbf{I}_{\mathrm{b}}(u,v)-\mu_{\mathrm{b}}\right|^2,
\end{equation}
where
\begin{equation}
	\mu_{\mathrm{b}} = \frac{1}{UV}\sum\limits_{u=1}^{U}\sum\limits_{v=1}^{V}\mathbf{I}_{\mathrm{b}}(u,v)
\end{equation}
is the mean of $\mathbf{I}_{\mathrm{b}}(u,v)$, is calculated. A lower variance corresponds to a homogeneous block and a higher variance to a heterogeneous block. For the calculation of the variance, the block $\mathbf{I}_{\mathrm{b}}(u,v)$ is enlarged to the block size plus $p$ pixels in order to have a more reliable calculation of the variance. Pixels from the loss area $\mathcal{B}$ are again neglected for the calculation.
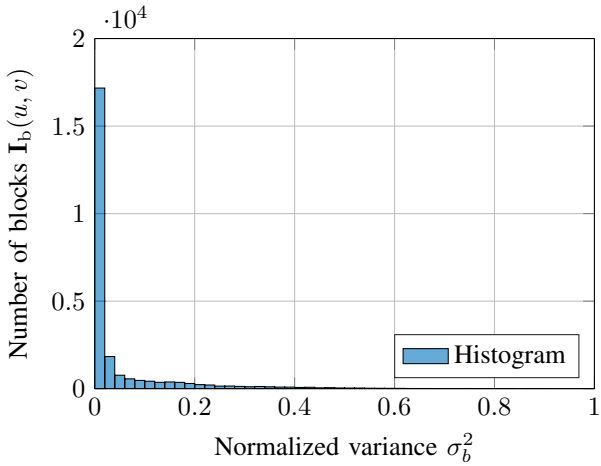
\begin{figure}
	\centering
%
%
\definecolor{mycolor1}{rgb}{0.00000,0.44700,0.74100}%
\definecolor{mycolor2}{rgb}{0.85000,0.32500,0.09800}%
\definecolor{mycolor3}{rgb}{0.00000,0.44706,0.74118}%
\definecolor{mycolor4}{rgb}{0.85098,0.32549,0.09804}%
\definecolor{mycolor5}{rgb}{0,0,0}%
\begin{tikzpicture}

\begin{axis}[%
width=0.75\columnwidth,
height=0.526\columnwidth,
at={(0\columnwidth,0\columnwidth)},
scale only axis,
xmin=0,
xmax=1,
xlabel={Normalized variance $\sigma^2_b$},
xmajorgrids,
separate axis lines,
every outer y axis line/.append style={mycolor5},
every y tick label/.append style={font=\color{mycolor5}},
ymin=0,
ymax=20000,
ylabel={Number of blocks $\mathbf{I}_{\mathrm{b}}(u,v)$},
ymajorgrids,
axis background/.style={fill=white},
yticklabel pos=left,
legend style={at={(0.97,0.03)},anchor=south east,legend cell align=left,align=left,draw=white!15!black}
]
\addplot[fill=mycolor1,fill opacity=0.6,draw=black,ybar interval,area legend] plot table[row sep=crcr] {%
x	y\\
0	17171\\
0.02	1836\\
0.04	769\\
0.06	562\\
0.08	473\\
0.1	426\\
0.12	363\\
0.14	384\\
0.16	362\\
0.18	297\\
0.2	233\\
0.22	209\\
0.24	147\\
0.26	151\\
0.28	130\\
0.3	111\\
0.32	126\\
0.34	107\\
0.36	91\\
0.38	89\\
0.4	71\\
0.42	80\\
0.44	48\\
0.46	58\\
0.48	41\\
0.5	37\\
0.52	36\\
0.54	27\\
0.56	24\\
0.58	21\\
0.6	18\\
0.62	12\\
0.64	10\\
0.66	10\\
0.68	8\\
0.7	7\\
0.72	6\\
0.74	10\\
0.76	1\\
0.78	5\\
0.8	1\\
0.82	2\\
0.84	0\\
0.86	2\\
0.88	1\\
0.9	0\\
0.92	0\\
0.94	0\\
0.96	0\\
0.98	3\\
1	3\\
};
\addlegendentry{Histogram};
\end{axis}
\end{tikzpicture}%
	\caption{A typical histogram of an image (\textit{Kodim19}) showing the number of blocks over the corresponding normalized variance.}
	\label{fig:hist_mapping}
\end{figure}
In Figure~\ref{fig:hist_mapping}, a typical histogram of the variances of different blocks $\mathbf{I}_{\mathrm{b}}(u,v)$ of an image is shown, where the variances of all blocks are normalized to values between $0$ and $1$. It can be seen that the number of homogeneous blocks (fewer variance) is higher than the number of heterogeneous blocks (higher variance).
Now, the mapping function
\begin{equation}
	i_{\mathrm{b,v}} = \left(i_{\mathrm{max}}-i_{\mathrm{min}}\right)\cdot \sigma^2_{\mathrm{b}} + \frac{1}{B}\left(i_{\mathrm{t,f}} - \sum\limits_{\tilde{b}=1}^{B}(i_{\mathrm{max}}-i_{\mathrm{min}})\cdot\sigma_{\mathrm{\tilde{b}}}^2\right)
\end{equation}
is applied in order to obtain the variable iterations $i_{\mathrm{b,v}}$ for each block depending on the corresponding variances $\sigma_{\mathrm{b}}^2$. In this paper, a linear mapping \mbox{$\left(i_{\mathrm{max}}-i_{\mathrm{min}}\right)\cdot \sigma_{\mathrm{b}}^2 + i_{\mathrm{min}}$} is used, however, other mapping functions are also possible.
On the one hand, the mapping function is employed to set a minimum number of iterations $i_{\mathrm{min}}$ to totally homogeneous blocks in order to obtain a reasonable reconstruction quality. 
On the other hand, it is used to spend more iterations of up to a maximum of $i_{\mathrm{max}}$ on heterogeneous blocks which require more iterations for visually pleasing results.
Depending on the content of the image to be reconstructed and the parameters of the mapping function, not all iterations $i_{\mathrm{t,f}}$ that the original FSR uses are distributed by \mbox{TD-FSR}. Therefore, the second term of $i_{\mathrm{b,v}}$ ensures that all remaining iterations are uniformly distributed over all blocks.
It follows that
\begin{equation}
	i_{\mathrm{t,v}} = \sum\limits_{b=1}^{B}i_{\mathrm{b,v}} = \sum\limits_{b=1}^{B}i_{\mathrm{b,f}} = i_{\mathrm{t,f}}.
\end{equation}
This means that \mbox{TD-FSR} spends the same amount of iterations $i_{\mathrm{t,v}}$ as the original FSR. They are, however, distributed depending on the texture of the blocks to be reconstructed.

\section{Simulations and Results}
\label{sec:results}
\begin{table}[]
	\caption{Parameters used by FSR and \mbox{TD-FSR}.}
	\label{tab:fsr_parameter}
	\centering	
	\begin{tabularx}{\columnwidth}{p{0.1cm}lc}
		\toprule
		 & Block size                                                    &  $4 \times 4$  \\
		 & Border width                                                  &      $14$      \\
		 & FFT size                                                      & $32 \times 32$ \\
		 & Average number of iterations per block $\bar{i}_{\mathrm{b}}$ & $20\dots 100$  \\
		 & Decay factor $\hat{\rho}$                                     &     $0.7$      \\
		 & Orthogonality deficiency compensation $\gamma$                &     $0.5$      \\
		 & Weighting of already reconstructed areas $\delta$             &     $0.5$      \\ \bottomrule
	\end{tabularx}
\end{table}
In the following, the performance of the proposed \mbox{TD-FSR} is evaluated. In general, \mbox{TD-FSR} and all other comparative algorithms may be applied on arbitrarily shaped loss regions as shown in~\cite{Schnurrer2015}, however, in this contribution only non-regularly sampled images are regarded. For simulation and evaluation, two image data sets are used, where the first one, the KODAK data base~\cite{Kodak2013}, consists of $24$ images of size \mbox{$768 \times 512$} pixels and is employed for the parameter training. It turned out that a surrounding of \mbox{$p=2$} should be taken for the calculation of the variance $\sigma_b^2$ and \mbox{$i_{\mathrm{min}}=10$} and \mbox{$i_{\mathrm{max}}=300$} as parameters for the mapping function. 
The second test set, the TECNICK data base~\cite{Asuni2011}, consists of $100$ images of size \mbox{$1200 \times 1200$} pixels and is employed for the actual simulation. Only the luminance component of the images is considered and all images are first multiplied with the same non-regular sampling mask, where $75\%$ of the pixels are missing and reconstructed afterwards using different algorithms. 

%

To determine the reconstruction quality, PSNR and SSIM~\cite{Wang2004} are calculated between the original image and the reconstructed image. For a comprehensive evaluation, \mbox{TD-FSR} is not only compared to FSR, but also to other algorithms that can be used for this reconstruction task. Namely, linear interpolation (LIN), steering kernel regression (SKR)~\cite{Takeda2007}, the constrained split augmented Lagrangian shrinkage algorithm (CLS) \cite{Afonso2011}, and sparsity-based wavelet inpainting (WI)~\cite{Starck2010}. For all mentioned algorithms, source code and parameters are taken from the literature. Both FSR and \mbox{TD-FSR} use the parameters shown in Table~\ref{tab:fsr_parameter}, where only the average number of iterations per block $\bar{i}_{\mathrm{b}}$ is varied from $20$ to $100$ in order to see the effect of the variable iterations of \mbox{TD-FSR}. This is denoted by $\text{FSR}_{\bar{i}_{\mathrm{b}}}$ and \mbox{$\text{TD-FSR}_{\bar{i}_{\mathrm{b}}}$}.
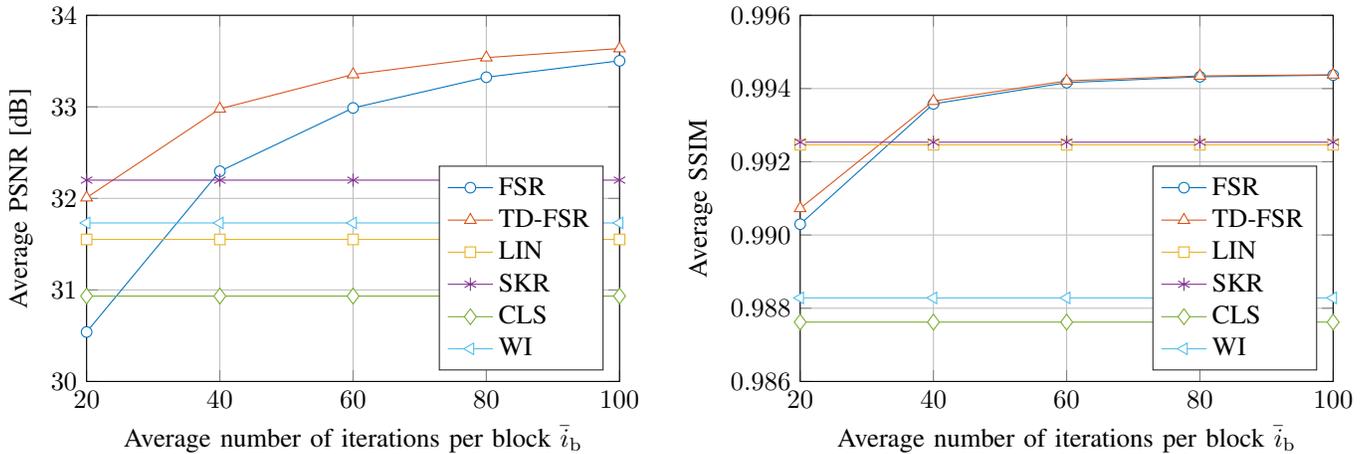
\begin{figure*}[t]
	\centering
%
%
\definecolor{mycolor1}{rgb}{0.00000,0.44700,0.74100}%
\definecolor{mycolor2}{rgb}{0.85000,0.32500,0.09800}%
\definecolor{mycolor3}{rgb}{0.92900,0.69400,0.12500}%
\definecolor{mycolor4}{rgb}{0.49400,0.18400,0.55600}%
\definecolor{mycolor5}{rgb}{0.46600,0.67400,0.18800}%
\definecolor{mycolor6}{rgb}{0.30100,0.74500,0.93300}%
\begin{tikzpicture}

\begin{axis}[%
width=0.8\columnwidth,
height=0.55\columnwidth,
at={(0\columnwidth,0\columnwidth)},
scale only axis,
xmin=20,
xmax=100,
xlabel={Average number of iterations per block $\bar{i}_{\mathrm{b}}$},
xmajorgrids,
ymin=30,
ymax=34,
ylabel={Average PSNR [dB]},
ymajorgrids,
axis background/.style={fill=white},
legend style={at={(0.97,0.03)},anchor=south east,legend cell align=left,align=left,draw=white!15!black}
]

\addplot [color=mycolor1,solid,mark size=2.0pt,mark=*,mark options={solid,fill=white}]
  table[row sep=crcr]{%
20	30.5403223894997\\
40	32.296603394915\\
60	32.9877823743573\\
80	33.3234764560415\\
100	33.5031939206745\\
};
\addlegendentry{FSR};

\addplot [color=mycolor2,solid,mark size=2.5pt,mark=triangle*,mark options={solid,fill=white}]
  table[row sep=crcr]{%
20	32.0080310612784\\
40	32.9784167592718\\
60	33.3535191452747\\
80	33.5380523562121\\
100	33.63712951962\\
};
\addlegendentry{TD-FSR};

\addplot [color=mycolor3,solid,mark size=2.0pt,mark=square*,mark options={solid,fill=white}]
  table[row sep=crcr]{%
20	31.5516110591992\\
40	31.5516110591992\\
60	31.5516110591992\\
80	31.5516110591992\\
100	31.5516110591992\\
};
\addlegendentry{LIN};

\addplot [color=mycolor4,solid,mark size=2.5pt,mark=asterisk,mark options={solid,fill=white}]
  table[row sep=crcr]{%
20	32.1993487203286\\
40	32.1993487203286\\
60	32.1993487203286\\
80	32.1993487203286\\
100	32.1993487203286\\
};
\addlegendentry{SKR};

\addplot [color=mycolor5,solid,mark size=3.0pt,mark=diamond*,mark options={solid,fill=white}]
  table[row sep=crcr]{%
20	30.9329203578057\\
40	30.9329203578057\\
60	30.9329203578057\\
80	30.9329203578057\\
100	30.9329203578057\\
};
\addlegendentry{CLS};

\addplot [color=mycolor6,solid,mark size=2.5pt,mark=triangle*,mark options={solid,rotate=90,fill=white}]
  table[row sep=crcr]{%
20	31.7317585614117\\
40	31.7317585614117\\
60	31.7317585614117\\
80	31.7317585614117\\
100	31.7317585614117\\
};
\addlegendentry{WI};

\end{axis}
\end{tikzpicture}%
\hspace*{0.3cm}
%
%
\definecolor{mycolor1}{rgb}{0.00000,0.44700,0.74100}%
\definecolor{mycolor2}{rgb}{0.85000,0.32500,0.09800}%
\definecolor{mycolor3}{rgb}{0.92900,0.69400,0.12500}%
\definecolor{mycolor4}{rgb}{0.49400,0.18400,0.55600}%
\definecolor{mycolor5}{rgb}{0.46600,0.67400,0.18800}%
\definecolor{mycolor6}{rgb}{0.30100,0.74500,0.93300}%
\begin{tikzpicture}

\begin{axis}[%
width=0.8\columnwidth,
height=0.55\columnwidth,
at={(0\columnwidth,0\columnwidth)},
scale only axis,
xmin=20,
xmax=100,
xlabel={Average number of iterations per block $\bar{i}_{\mathrm{b}}$},
xmajorgrids,
ymin=0.986,
ymax=0.996,
yticklabel style={
            /pgf/number format/fixed,
            /pgf/number format/precision=3,
            /pgf/number format/fixed zerofill
        },
        scaled y ticks=false,
ylabel={Average SSIM},
ymajorgrids,
axis background/.style={fill=white},
legend style={at={(0.97,0.03)},anchor=south east,legend cell align=left,align=left,draw=white!15!black}
]

\addplot [color=mycolor1,solid,mark size=2.0pt,mark=*,mark options={solid,fill=white}]
  table[row sep=crcr]{%
20	0.990295635126771\\
40	0.99358123683135\\
60	0.994161139574324\\
80	0.994321193271385\\
100	0.994365815721756\\
250	0.994260250268976\\
};
\addlegendentry{FSR};

\addplot [color=mycolor2,solid,mark size=2.5pt,mark=triangle*,mark options={solid,fill=white}]
  table[row sep=crcr]{%
20	0.990728224814397\\
40	0.993656167880628\\
60	0.994210404858843\\
80	0.994345313524393\\
100	0.994376726422871\\
250	0.99426068582309\\
};
\addlegendentry{TD-FSR};

\addplot [color=mycolor3,solid,mark size=2.0pt,mark=square*,mark options={solid,fill=white}]
  table[row sep=crcr]{%
20	0.992463372981107\\
40	0.992463372981107\\
60	0.992463372981107\\
80	0.992463372981107\\
100	0.992463372981107\\
250	0.992463372981107\\
};
\addlegendentry{LIN};

\addplot [color=mycolor4,solid,mark size=2.5pt,mark=asterisk,mark options={solid,fill=white}]
  table[row sep=crcr]{%
20	0.992534843910879\\
40	0.992534843910879\\
60	0.992534843910879\\
80	0.992534843910879\\
100	0.992534843910879\\
250	0.992534843910879\\
};
\addlegendentry{SKR};

\addplot [color=mycolor5,solid,mark size=3.0pt,mark=diamond*,mark options={solid,fill=white}]
  table[row sep=crcr]{%
20	0.987620966474121\\
40	0.987620966474121\\
60	0.987620966474121\\
80	0.987620966474121\\
100	0.987620966474121\\
250	0.987620966474121\\
};
\addlegendentry{CLS};

\addplot [color=mycolor6,solid,mark size=2.5pt,mark=triangle*,mark options={solid,rotate=90,fill=white}]
  table[row sep=crcr]{%
20	0.988282156898214\\
40	0.988282156898214\\
60	0.988282156898214\\
80	0.988282156898214\\
100	0.988282156898214\\
250	0.988282156898214\\
};
\addlegendentry{WI};

\end{axis}
\end{tikzpicture}%
	\caption{Average PSNR and average SSIM results for the proposed TD-FSR compared to different comparative reconstruction algorithms plotted over the average number of iterations per block $\bar{i}_{\mathrm{b}}$ and evaluated on the TECNICK data base.}
	\label{fig:results_tecnick_psnr_over_iterations}
\end{figure*}

The PSNR and SSIM results of the simulations are shown in Figure~\ref{fig:results_tecnick_psnr_over_iterations}.
In the left graph, it can be seen that for $20$ iterations, TD-FSR achieves a PSNR gain of $1.47$~dB compared to FSR and a better reconstruction quality than LIN, CLS, and WI. Despite the fact that $\text{TD-FSR}_{20}$ gives a significantly better reconstruction quality than most of the other algorithms, it is slightly below SKR. However, the reconstruction quality of both \mbox{FSR} and \mbox{TD-FSR} gets better with an increasing number of iterations. With at least $40$ iterations both algorithms give better results than LIN, SKR, CLS, and WI. Additionally, $\text{TD-FSR}_{40}$ has a PSNR gain of $0.68$~dB compared to $\text{FSR}_{40}$. For a higher number of iterations, however, the reconstruction quality of \mbox{TD-FSR} is only slightly better than the reconstruction quality of FSR. This means that if a very high number of iterations is spent, both algorithms should converge to the same reconstruction quality. Compared to the best of the other comparative algorithms, namely SKR, $\text{TD-FSR}_{100}$ can achieved a gain in PSNR of up to $1.44$~dB. As a second quality measure, SSIM values are displayed in the right graph of Figure~\ref{fig:results_tecnick_psnr_over_iterations}. It can be seen that SSIM shows almost the same behavior as PSNR. It shows the gains in reconstruction quality of $\text{TD-FSR}$ compared to $\text{FSR}$ especially for a small number of iterations and a similar quality for a higher number of iterations. 

For a more precise statement regarding the complexity of the different algorithms, the average reconstruction time for one image of the TECNICK data base is measured. The tests are carried out on an Intel Core i7-3770~ @~$3.40$~GHz, equipped with $32$~GB of RAM and running MATLAB R2015b.
\setlength{\tabcolsep}{0.29cm}
\begin{table}[]
	\caption{Average processing time in seconds for one image of the TECNICK data base.}
	\label{tab:runtime_results}
	\centering	
	\begin{tabularx}{\columnwidth}{cccccc}
		  $\text{TD-FSR}_{20\ldots 100}$ & $\text{FSR}_{20\ldots 100}$ &  LIN   &  SKR  & CLS  &   WI   \\ \midrule
		          $98\ldots 333$         &       $96\ldots 331$        & $7.60$ & $189$ & $70$ & $9565$
	\end{tabularx}
\end{table}
The results can be seen in Table~\ref{tab:runtime_results}. LIN is a very fast reconstruction algorithm ($7.60$~s), whereas WI requires a lot of processing time ($9565$~s). SKR ($189$~s) and CLS ($70$~s) are algorithms with moderate speed. Using \mbox{FSR} ($96\ldots 331$~s) or \mbox{TD-FSR} ($98\ldots 333$~s), the processing time depends on the number of iterations, but it is still moderate.

\begin{figure*}[ht]
	\centering
	\def\svgwidth{\textwidth}
	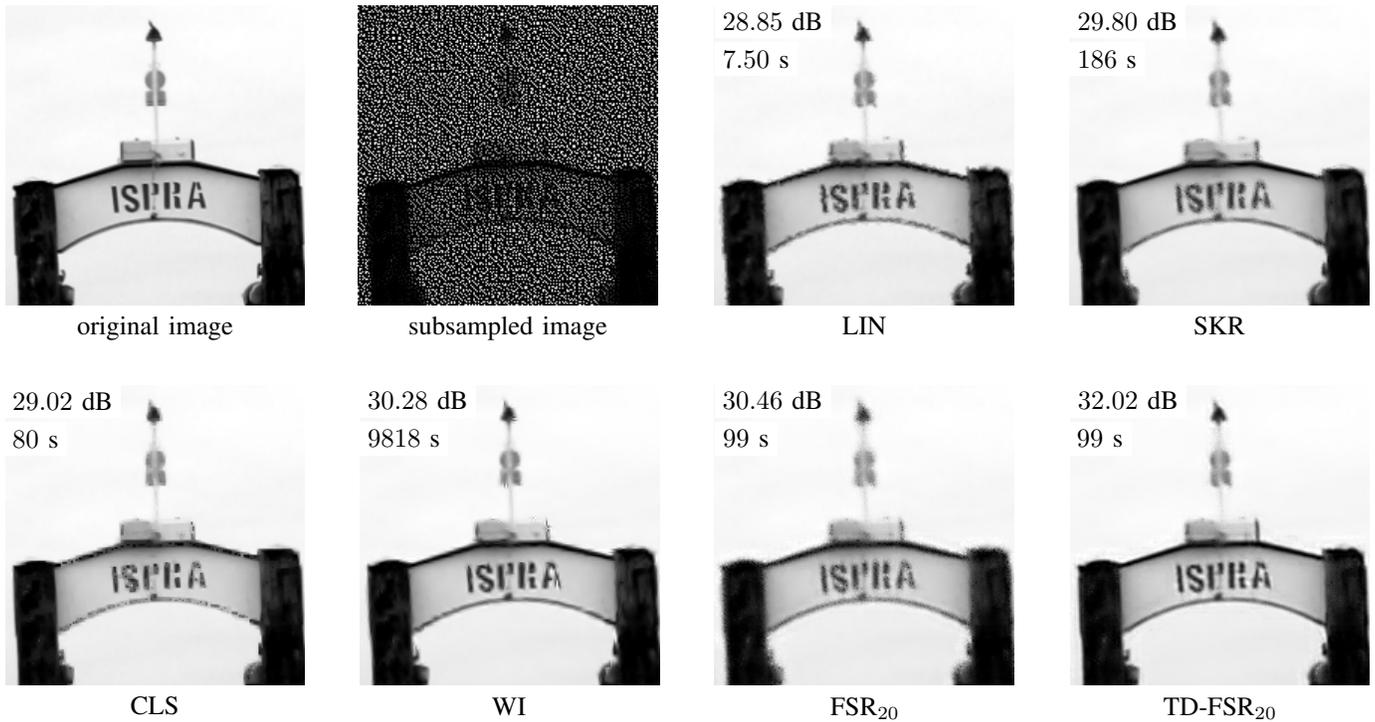
	\caption{Image detail example for a visual comparison of the proposed \mbox{TD-FSR} with other comparative reconstruction algorithms including PSNR and runtime values (measured on the entire image).}
	\label{fig:image_examples}
	\vspace*{-0.2cm}
\end{figure*}

In Figure~\ref{fig:image_examples}, an image detail example is given for a visual comparison of the proposed \mbox{TD-FSR} with the other comparative reconstruction algorithms.
The given PSNR values and processing times are measured on the entire image.
It can be seen that LIN and CLS fail to reconstruct sharp edges or fine details. SKR and WI are able to preserve sharp edges, but introduce other artifacts. FSR with only $20$ iterations gives a higher PSNR value than LIN, SKR, CLS, and WI, yielding, however, a similar visual quality which is still improvable. With the proposed \mbox{TD-FSR} and also only $20$ iterations the best PSNR value is obtained and a high visual quality can be achieved where fine details and sharp edges are reconstructed. The processing time of \mbox{$\text{TD-FSR}_{20}$} is faster than WI and SKR and comparable to CLS. Only LIN is faster, yielding, however, a very poor reconstruction quality. 

The evaluations show, that compared to other \mbox{state-of-the-art} reconstruction methods, the proposed \mbox{TD-FSR} gives the best reconstruction quality in terms of PSNR, SSIM and also visually. Compared to the original \mbox{FSR}, the highest gains in PSNR and SSIM can be achieved for a small number of iterations. This means, that \mbox{TD-FSR} is especially suitable in scenarios where only little computational power is available, for example in mobile devices. It can also be stated that the additional processing time of \mbox{TD-FSR} that is introduced due to the image segmentation and distribution of the iterations is negligible.

\section{Conclusion}
\label{sec:conclusion}
In this paper, a new texture-dependent approach for frequency selective reconstruction has been introduced. Frequency selective reconstruction is used for many signal reconstruction tasks like image restoration or object removal, but it can also be applied for the reconstruction of non-regularly sampled images. Since frequency selective reconstruction is an iterative algorithm and it has been shown that it is not required to spend the same amount of iterations to both homogeneous and heterogeneous regions, it has been proposed to distribute the number of iterations depending on the texture of the blocks that have to be reconstructed. Compared to other \mbox{state-of-the-art} reconstruction algorithms, this leads to a visually noticeable gain in PSNR of up to $1.44$~dB and compared to the original frequency selective reconstruction, gains of up to $1.47$~dB depending on the number of iterations can be achieved.

Future work might extend the texture-dependent approach to the 3-dimensional frequency selective reconstruction. Since many more iterations are required for the 3-dimensional sparse model generation, more gains from a sophisticated distribution of the number of iterations per cube are expected.

\section*{Acknowledgment}
This work has been supported by the Deutsche Forschungsgemeinschaft (DFG) under contract number KA 926/5-3.

\bibliographystyle{IEEEtran}
\bibliography{bib_strings_long,literature}

\begin{thebibliography}{10}
\providecommand{\url}[1]{#1}
\csname url@samestyle\endcsname
\providecommand{\newblock}{\relax}
\providecommand{\bibinfo}[2]{#2}
\providecommand{\BIBentrySTDinterwordspacing}{\spaceskip=0pt\relax}
\providecommand{\BIBentryALTinterwordstretchfactor}{4}
\providecommand{\BIBentryALTinterwordspacing}{\spaceskip=\fontdimen2\font plus
\BIBentryALTinterwordstretchfactor\fontdimen3\font minus
  \fontdimen4\font\relax}
\providecommand{\BIBforeignlanguage}[2]{{%
\expandafter\ifx\csname l@#1\endcsname\relax
\typeout{** WARNING: IEEEtran.bst: No hyphenation pattern has been}%
\typeout{** loaded for the language `#1'. Using the pattern for}%
\typeout{** the default language instead.}%
\else
\language=\csname l@#1\endcsname
\fi
#2}}
\providecommand{\BIBdecl}{\relax}
\BIBdecl

\bibitem{Duparre2006}
J.~W. Duparré and F.~C. Wippermann, ``{M}icro-optical {A}rtificial {C}ompound
  {E}yes,'' \emph{Bioinspiration and Biomimetics}, vol.~1, no.~1, pp. R1--R16,
  Apr. 2006.

\bibitem{Hennenfent2007}
G.~Hennenfent and F.~J. Herrmann, ``{I}rregular {S}ampling: {F}rom {A}liasing
  to {N}oise,'' in \emph{Proc. 69th EAGE Conference \& Exhibition Incorporating
  SPE EUROPEC}, London, United Kingdom, Jun. 2007.

\bibitem{Schoeberl2011}
M.~Schöberl, J.~Seiler, S.~Fößel, and A.~Kaup, ``{I}ncreasing {I}maging
  {R}esolution by {C}overing {Y}our {S}ensor,'' in \emph{Proceedings {IEEE}
  International Conference on Image Processing ({ICIP})}, Brussels, Belgium,
  Sep. 2011, pp. 1937--1940.

\bibitem{Seiler2015}
J.~Seiler, M.~Jonscher, M.~Schöberl, and A.~Kaup, ``{R}esampling {I}mages to a
  {R}egular {G}rid from a {N}on-{R}egular {S}ubset of {P}ixel {P}ositions
  {U}sing {F}requency {S}elective {R}econstruction,'' \emph{{IEEE} Transactions
  on Image Processing}, vol.~24, no.~11, pp. 4540--4555, Nov. 2015.

\bibitem{Takeda2007}
H.~Takeda, S.~Farsiu, and P.~Milanfar, ``{K}ernel {R}egression for {I}mage
  {P}rocessing and {R}econstruction,'' \emph{{IEEE} Transactions on Image
  Processing}, vol.~16, no.~2, pp. 349--366, Feb. 2007.

\bibitem{Afonso2011}
M.~V. Afonso, J.~M. Bioucas-Dias, and M.~A.~T. Figueiredo, ``{A}n {A}ugmented
  {L}agrangian {A}pproach to the {C}onstrained {O}ptimization {F}ormulation of
  {I}maging {I}nverse {P}roblems,'' \emph{{IEEE} Transactions on Image
  Processing}, vol.~20, no.~3, pp. 681--695, Mar. 2011.

\bibitem{Starck2010}
J.-L. Starck, F.~Murtagh, and J.~Fadili, \emph{{S}parse {I}mage and {S}ignal
  {P}rocessing: {W}avelets, {C}urvelets, {M}orphological {D}iversity}.\hskip
  1em plus 0.5em minus 0.4em\relax New York, NY, USA: Cambridge University
  Press, 2010.

\bibitem{Seiler2010}
J.~Seiler and A.~Kaup, ``{C}omplex-{V}alued {F}requency {S}elective
  {E}xtrapolation for {F}ast {I}mage and {V}ideo {S}ignal {E}xtrapolation,''
  \emph{{IEEE} Signal Processing Letters}, vol.~17, no.~11, pp. 949--952, Nov.
  2010.

\bibitem{Koloda2014}
J.~Koloda, J.~Seiler, A.~Kaup, V.~Sánchez, and A.~M. Peinado, ``{A}n
  {E}rror-{B}ased {R}ecursive {F}illing {O}rdering for {I}mage {E}rror
  {C}oncealment,'' in \emph{Proceedings {IEEE} International Conference on
  Image Processing ({ICIP})}, Paris, France, Oct. 2014, pp. 2517--2521.

\bibitem{Jonscher2016}
M.~Jonscher, K.~Jaskolka, J.~Seiler, and A.~Kaup, ``{R}ecursive {F}requency
  {S}elective {R}econstruction of {N}on-{R}egularly {S}ampled {V}ideo {D}ata,''
  in \emph{submitted to Proceedings Picture Coding Symposium ({PCS})},
  Nuremberg, Germany, Dec. 2016.

\bibitem{Schnurrer2015}
W.~Schnurrer, M.~Jonscher, J.~Seiler, T.~Richter, M.~Bätz, and A.~Kaup,
  ``{C}entroid {A}dapted {F}requency {S}elective {E}xtrapolation for
  {R}econstruction of {L}ost {I}mage {A}reas,'' in \emph{Proceedings {IEEE}
  Visual Communications and Image Processing ({VCIP})}, Singapore, Singapore,
  Dec. 2015, pp. 1--4.

\bibitem{Kodak2013}
``{K}odak {T}est {I}mage {L}ibrary,'' Jan. 2013, http://r0k.us/graphics/kodak/.

\bibitem{Asuni2011}
N.~Asuni, ``{T}ecnick {T}est {I}mage {L}ibrary,'' Apr. 2011,
  http://testimages.tecnick.com.

\bibitem{Wang2004}
Z.~Wang, A.~Bovik, H.~Sheikh, and E.~Simoncelli, ``{I}mage {Q}uality
  {A}ssessment: {F}rom {E}rror {V}isibility to {S}tructural {S}imilarity,''
  \emph{{IEEE} Transactions on Image Processing}, vol.~13, no.~4, pp. 600--612,
  Apr. 2004.

\end{thebibliography}
\label{sec:ref}

\end{document}